\documentclass{IEEEojcsys}

\usepackage[colorlinks,urlcolor=blue,linkcolor=blue,citecolor=blue]{hyperref}

\usepackage[utf8]{inputenc} % allow utf-8 input
\usepackage[T1]{fontenc}    % use 8-bit T1 fonts
\usepackage{url}            % simple URL typesetting
\usepackage{booktabs}       % professional-quality tables
\usepackage{amsfonts}       % blackboard math symbols
\usepackage{nicefrac}       % compact symbols for 1/2, etc.
\usepackage{microtype}      % microtypography
\usepackage{xcolor}         % colors

\usepackage{multirow}
\usepackage{threeparttable}

\jvol{00}
\jnum{XX}
\paper{1234567}
\pubyear{2024}
\receiveddate{01 May 2024}
\reviseddate{17 August 2024}
\accepteddate{19 August 2024}
\currentdate{20 Aug 2024}
\usepackage{graphicx} 

\usepackage{graphicx} 
\usepackage{algorithm}
\usepackage{algpseudocode}
\usepackage{amsmath}
\usepackage{amssymb}
\usepackage{amsthm}

\theoremstyle{definition}

\usepackage{bbm}
\usepackage{todonotes}
\usepackage{soul}

\usepackage{subfigure}
\usepackage{microtype,wrapfig}
\usepackage{cite} %for citing range of papers [1-5]

\usepackage[english]{babel}
\usepackage[numbers]{natbib}

\algnewcommand\algorithmicforeach{\textbf{for each}}
\algdef{S}[FOR]{ForEach}[1]{\algorithmicforeach\ #1\ \algorithmicdo}

\DeclareMathOperator*{\E}{\mathbb{E}}

\renewcommand{\thesection}{\Roman{section}}
\renewcommand{\thesubsection}{\Alph{subsection}}

\makeatletter
\renewcommand{\p@subsection}{\thesection.}
\renewcommand{\p@subsubsection}{\thesection.\thesubsection.}
\makeatother

\begin{document}

\sptitle{Article Category}

\title{Hamilton-Jacobi Reachability in Reinforcement~Learning: A Survey} 

\editor{This paper was recommended by Associate Editor Peter Seiler.}

\author{Milan Ganai\affilmark{1}}

\author{Sicun Gao\affilmark{1} (Member, IEEE)}

\author{Sylvia Herbert\affilmark{2} (Member, IEEE)}

\affil{Department of Computer Science and Engineering, University of California San Diego, La Jolla, CA 92093 USA} 
\affil{Department of Mechanical and Aerospace Engineering, University of California San Diego, La Jolla, CA 92093 USA}

\corresp{CORRESPONDING AUTHOR: MILAN GANAI (e-mail: \href{mailto:mganai@ucsd.edu}{mganai@ucsd.edu})}
\authornote{This material is based on the work supported by ONR YIP N00014-22-1-2292, NSF Career CCF 2047034, NSF CCF DASS 2217723, and NSF AI Institute CCF 2112665.}

\markboth{HAMILTON-JACOBI REACHABILITY IN REINFORCEMENT LEARNING: A SURVEY}{GANAI {\itshape ET AL}.}

\begin{abstract}
Recent literature has proposed approaches that learn control policies with high performance while maintaining safety guarantees. Synthesizing Hamilton-Jacobi (HJ) reachable sets has become an effective tool for verifying safety and supervising the training of reinforcement learning-based control policies for complex, high-dimensional systems. Previously, HJ reachability was restricted to verifying low-dimensional dynamical systems primarily because the computational complexity of the dynamic programming approach it relied on grows exponentially with the number of system states. In recent years, a litany of proposed methods addresses this limitation by computing the reachability value function simultaneously with learning control policies to scale HJ reachability analysis while still maintaining a reliable estimate of the true reachable set. These HJ reachability approximations are used to improve the safety, and even reward performance, of learned control policies and can solve challenging tasks such as those with dynamic obstacles and/or with lidar-based or vision-based observations. In this survey paper, we review the recent developments in the field of HJ reachability estimation in reinforcement learning that would provide a foundational basis for further research into reliability in high-dimensional systems.
\end{abstract}


\maketitle

\section{Introduction}

As autonomous control systems are deployed in the real world, there is a growing need to develop methods with rigorous safety guarantees to avert failure in critical decision points, mitigate risk of unpredictability, and safeguard users' trust in the system. Verification-based approaches relying on control theoretic functions have been in the forefront among studied solutions. However, the large uncertainty and complex nature of real world dynamics limits the practical application of many of these approaches. 

Hamilton-Jacobi (HJ) reachability analysis is a rigorous tool that verifies the safety and/or liveness of a dynamic system~\cite{chen2018hamilton, hjreachabilityoverview}. For a specified model and target set, HJ reachability analysis is typically used to compute the set of initial states from which the system can reach a goal despite bounded disturbance. For safety analysis, HJ reachability can provide the set of initial states from which the system may be forced into the failure set despite best-case efforts (the complement of this set of initial states is, therefore, the safe set). This verification method provides guarantees on the safety properties of a system and the approach generalizes to various difficult problem settings. These include problems with nonlinear dynamics, reach-avoid problems with time-varying goals or constraints~\cite{fisac2015reach}, problems that must be robust to bounded system uncertainties or disturbances~\cite{chen2017reachability, chen2018robust}, and finding other certificate functions~\cite{gong2022constructing}. 

HJ reachability computation is based on finding a viscosity solution for the Hamilton-Jacobi-Bellman partial differential equation (HJBPDE) corresponding to a specified dynamics model and target set. Proposed approaches have accomplished this by discretizing the state space and using dynamic programming mechanisms~\cite{bellman1966dynamic}. However, this approach has been practically deployed on systems with at most 6 dimensions~\cite{decomposition, bui2022optimizeddp}. The main challenge is that the computational complexity of these approaches is exponential in the state dimensions~\cite{hjreachabilityoverview}, rendering them intractable in relatively large dimension systems.

To address this issue on the curse of dimensionality, past works have proposed approaches that make strong assumptions such as convexity, order preserving dynamics, and mixed monotone systems~\cite{darbon2016algorithms, hafner2009computation, coogan2015efficient} or exploit the system's structure~\cite{fisac2015reach, mitchell2011scalable, mitchell2003overapproximating, chen2016fast, kaynama2009schur, kaynama2013modified}. However, these approaches still do not necessarily scale well with the complexity encountered in the learning-based controls. Furthermore, they still require access to the model for active sampling and/or computation of gradients of the dynamics.

In this survey, we focus on a recent line of work that learns the HJ reachability value function in conjunction with learning control policies. Particularly, recent approaches like~\cite{akametalu2018minimum, fisac2019bridging} demonstrated how to learn a discrete-time value function solution of the HJBPDE via a recursive Bellman formulation. These value functions describe the maximum reachability violation or reward (depending on the usage) that a particular control policy achieves from each state. This form of learning has opened a new direction of research in which the learned reachability value function can directly be incorporated in reach-avoid problems~\cite{hsu2021safety} and safety-constrained reinforcement learning~\cite{yu2022reachability, ganai2023iterative}. While learning a certificate has been implemented for other safety verification functions (e.g. control barrier functions), significant benefits of learning reachability value functions include the ability to guarantee convergence to a valid solution of the HJBPDE of a particular control policies' dynamics. Learned reachability value functions for learned control policies have been demonstrated to be effective in various challenging problems~\cite{fisac2019bridging, hsu2021safety, yu2022reachability, ganai2023iterative, hsunguyen2023isaacs}.

\begin{table*}[t]
\centering
\caption{Classification of the primary works we discuss in this survey by control/safety problem addressed, model access, types of noise/disturbance handled, and the highest dimension state space on which results are published.}
\begin{threeparttable}
\begin{tabular}{ccccc}
\hline
\textit{Approach (Published Year)} & \textit{Problem Type} & \textit{Model Access} & \textit{Noise/Disturbance Considered} & \textit{Max State Dim.}  \\ \hline
\citet{hjreachabilityoverview} (2017) & Optimal Control & Model-based & Adversarial Disturbance & $10$D \\ \hline
\citet{akametalu2014reachability} (2014) & Safe RL & Model-based & Adversarial Disturbance & $4$D \\ 
\citet{fisac2018general} (2018) & Safe RL & Model-based & Adversarial Disturbance & $2$D \\ 
\citet{ivanovic2019barc} (2019) & Safe RL & Model-based &  Stochastic RL, Deterministic HJ & $6$D \\ \hline
\citet{akametalu2018minimum} (2018) & Optimal Control & Model-based & Adversarial Disturbance & $3$D  \\ 
\citet{fisac2019bridging} (2019) & Optimal Control & Model-free & None/Deterministic & $18$D \\ \hline
\citet{fisac2015reach} (2015) & Reach-Avoid & Model-based & Adversarial Disturbance & $3$D \\ 
\citet{hsu2021safety} (2021) & Reach-Avoid & Model-free & None/Deterministic & $6$D \\ 
\citet{so2023solving} (2023) & Stabilize-Avoid & Model-free & None/Deterministic & $17$D \\ \hline
\citet{chen2021safe} (2021) & Safe RL & Model-free & None/Deterministic & $40$D\tnote{a} \\ 
\citet{yu2022reachability} (2022) & Safe RL & Model-free & None/Deterministic & $112$D\tnote{b} \\ 
\citet{ganai2023iterative} (2023) & Safe RL & Model-free & Stochastic Dynamics & $76$D\tnote{b} \\ \hline
\citet{hsuren2022slr} (2022) & Robust Deployment & Model-free & Unseen/Random Environment & $90\times160$ pixel RGB\tnote{c} \\ 
\citet{hsunguyen2023isaacs} (2023) & Robust Deployment & Model-based & Adversarial \& Stochastic Disturbance & $5$D \\ 
\hline
\end{tabular}
\label{tab:ClassifySurvey}
\begin{tablenotes}
\item[a] Encodes $192\times144$ RGB ego-camera view and speed into $40$D state representation for actors and critics.
\item[b] Lidar-based state space.
\item[c] Actors and critics receive $4$ of these RGB images along with $10$D of latent variable and $2$D of auxiliary signal information.
\end{tablenotes}
\end{threeparttable}
\end{table*}

\subsection{Related Surveys}

While there are several recent surveys on related topics, none discuss the rapidly growing literature on HJ reachability for learned controls.~\citet{hjreachabilityoverview} reviews HJ reachability methods for high-dimensional reachability analysis (examples shown up to 10D) and includes a brief discussion on reachability analysis that use neural networks to solve HJBPDEs. Nonetheless, the approaches presented in the survey may not necessarily scale to the complexity encountered in systems controlled primarily with learned-based policies ($>$20D).~\citet{chen2018hamilton} presents approaches to scale HJ reachability verification through system decomposition of nonlinear dynamics and applications in unmanned airspace management, but does not discuss learning-based HJ reachability techniques. The 2021 survey by~\citet{althoff2021set} covers methods that find a guaranteed overapproximation of the reachability set via set propagation; however, it leaves to future work HJ reachability methods for online verification of partially known environments, as well as systems involving neural networks (note that we use the term online in this survey to mean the framework of actively interacting with an environment to acquire the optimal control policy). The recent survey by~\citet{dawson2023safe} covers topics on neural control certificates -- this class includes learning-based Lyapunov and Barrier functions~\cite{NLC, Chang2021, ganai2023learning, qin2022quantifying}. 

\subsection{Motivation and Challenges}

HJ reachability is a powerful tool in achieving safe and optimal control objectives across a variety of complex domains. However, its application in real world systems faces significant challenges that must be addressed to fully unlock its potential. 

HJ reachability can rigorously guarantee safety in dynamical systems by determining the set of states from which the system can be steered to a safe state under all possible disturbances. However, ensuring these guarantees are satisfied requires scrupulous consideration of the system's dynamics and constraints. Scalability is a central challenge in HJ reachability: as the state space of the system increases, the computational cost for solving the HJBPDE grows exponentially~\cite{hjreachabilityoverview}. Developing scalable algorithms to manage high-dimensional state spaces without sacrificing the accuracy of the reachability analysis is critical for extending HJ reachability to more complex systems such as those we discuss in Section~\ref{sec:BroaderApp}. Furthermore, it is important to adapt the methods in order to scalably solve the useful variants of HJ reachability: forward and backward reachability, as well as combining goal achievement (liveness) with danger avoidance (safety).

In various settings, direct access to the system's dynamics is unavailable, either due to incomplete knowledge of the system or because the system is too complex to model accurately. Integrating HJ reachability into this scenario requires innovative approaches~\cite{akametalu2014reachability} that leverage data-driven methods to interact with and learn from a Markov decision process interface (see Section~\ref{subsec:MDP}). A concomitant challenge is verifying these approximations do not compromise the safety guarantees provided by HJ reachability. Reinforcement learning (RL) offers a promising avenue for applying HJ reachability in scenarios where explicit system models are unavailable~\cite{fisac2019bridging}. Nonetheless, we must verify the solutions obtained via RL form valid viscosity solutions. 

In other practical applications, systems must often satisfy both hard constraints (those that cannot be violated) and soft constraints (those can be violated only to prioritize hard constraints). Traditionally, HJ reachability addresses hard constraint satisfaction, but integrating soft constraints into this framework requires novel methodologies that can balance these various constraint types while maintaining overall system safety~\cite{ganai2023iterative}.

HJ reachability is designed to handle worst-case disturbances, but this can lead to excessively conservative solutions. A more reasonable setting is stochastic dynamics, in which it is desirable to leverage the safety guarantees of HJ reachability without always having to anticipate the worst case~\cite{abate2008probabilistic}. Developing methods that incorporate probabilistic models while still providing useful safety guarantees is an ongoing area of research. Furthermore, another challenge is integrating HJ reachability with other certificate functions, such as control barrier and lyapunov functions~\cite{dawson2023safe}, which can help systems reach goals and return to safety after a violation. Additionally, in the context of continual lifelong learning~\cite{parisi2019continual}, it is important to allow HJ reachability methods to adapt as the system learns and evolves over time as well as maintain safety during the training procedure when interacting with the environment. All these challenges require scalable methods that can handle large-scale learning and dynamic updates for closed form solutions.

The Hopf formulation is efficient in solving HJ equations in linear dynamics~\cite{rublev2000generalized}, offering a potential direction to accelerate solution acquisition. Integrating this method into learning-based frameworks could significantly enhance both the scalability and performance of HJ reachability. Thus, another challenge is determining how the Hopf formulation can be combined with modern machine learning techniques.

In summary, while HJ reachability provides a powerful framework to guarantee system safety and achieve control objectives, its practical application faces a variety of challenges. In this survey, we examine the current progress in addressing these challenges: we discuss the development of novel methods that have enhanced scalability of HJ reachablity, its integration with reinforcement learning, and how it is employed to balance constraints and leverage the strengths of other certificate functions. We will also discuss what challenges still remain unresolved and future directions to investigate to rectify them.

\subsection{Broader Applications}
\label{sec:BroaderApp}

Dynamical systems are central in many fields, making obtaining optimal control integral for understanding these systems. The HJBPDE offers a robust framework to achieve this purpose: numerous applications have successfully reformulated their problems to fit the HJBPDE framework, highlighting the significant potential of HJ reachability estimation in various domains.

In the context of robotics, HJ reachability estimation has been employed to tackle optimal control problems in dynamical systems facing (adversarial) disturbances as well as addressing robustness problems. Some applications include controlling UAV drones in the presence of bounded-strength winds~\cite{fisac2018general} and acquiring safe policies in deploying quadruped robots~\cite{hsuren2022slr}. This progress lays the groundwork for future applications including humanoid robotics in both domestic and industrial environments. Some additional notable autonomous system applications of HJ reachability estimation include safe and stable control of F16 fighter jets~\cite{so2023solving}, race car control with vision (both image-based and lidar-based) input~\cite{chen2021safe}, and fuel-efficient navigation of spacecrafts~\cite{holzinger2011optimal}.

HJ reachability also has much potential in studying biological processes.~\citet{sharpless2023koopman} uses a HJ-based method to solve for optimal control to drive the biochemical process of yeast glycolysis toward some target ATP synthesis that a bioengineer may intend for cell growth. In the work of~\citet{gandon2017hamilton}, the authors propose analyzing evolutionary biology, particularly processes concerning genetic population distributions affected by mutation, selection, and migrations, with HJ methods.~\citet{padovano2024development} models the evolutionary dynamics of metastatic tumors under chemotherapy with HJB equations, paving way for advancements in accelerating drug discovery. Much progress can be achieved in studying high-dimensional biological processes with HJ reachability estimation techniques.

Energy generation and management applications have been analyzed through the framework of HJ reachability. For example, the tokamak, which is a device generating strong magnetic fields to restrict plasma in a toroidal shape~\cite{wesson2011tokamaks}, has been notably examined for its potential in fusion-based energy production. Studies like \citet{mcgann2010hamilton} demonstrate that HJ equations can model and control the magnetic field dynamics across the toroidal surface. Furthermore, \citet{heymann2016stochastic} addresses the issue of microgrid energy management by reformulating it as an HJB equation, employing real-world data from Chile. HJ reachability estimation methods have significant promise in delivering safe and efficient methods to address the global energy crisis~\cite{farghali2023strategies}.

Another novel application is generative modeling (GM): the work of~\citet{berneroptimal} which connects HJB equations to the stochastic differential equation for diffusion-based GM. They also demonstrate how to migrate methods from HJ-based optimal control theory to GM. Ultimately, with the increasing presence of dynamical system problems within generative AI, HJ reachability estimation methods have much potential to fundamentally understand and accelerate advancements in large scale sampling for GM.

\subsection{Survey Overview}
In this review we aim to provide an overview of estimating (i.e. via data-driven methods) HJ reachability specifically for learned controls. We provide a summary of the classification of the main papers that we discuss in this survey in Table~\ref{tab:ClassifySurvey}. We structure this survey in the following manner:
\begin{itemize}
    \item In Section~\ref{sec:Prelim}, we formally introduce reinforcement learning and HJ reachability analysis.
    \item In Section~\ref{sec:TradHJ_RL}, we discuss approaches that use traditional HJ reachability for learned control.
    \item In Section~\ref{sec:LearnHJ}, we demonstrate how to learn HJ reachability online to acquire reinforcement learning-based control.
    \item In Section~\ref{sec:ReachAvoid}, we survey various HJ reachability-based/-inspired methods that solve reach-avoid tasks.
    \item In Section~\ref{sec:ModelFreeSafeRL}, we review approaches for model-free safe reinforcement learning in both deterministic and stochastic dynamics scenarios.
    \item In Section~\ref{sec:RobustRW}, we examine HJ reachability estimation-based methods that address robustness and uncertainty issues found in real world environments.
    \item In Section~\ref{sec:Limitations}, we discuss the limitations of HJ reachability estimation approaches.
    \item In Section~\ref{sec:FutureWork}, we lay out new research directions for future works in using HJ reachability estimation.
\end{itemize}

\section{Preliminaries}
\label{sec:Prelim}
\subsection{Markov Decision Processes}
\label{subsec:MDP}

A Markov decision process (MDP) is defined as $\mathcal{M}:=\langle \mathcal{S}, \mathcal{A}, P, r, \gamma\rangle$, where
\begin{itemize}
    \item $\mathcal{S}\subseteq \mathbb{R}^n$ and $\mathcal{A}\subseteq \mathbb{R}^{m_a}$ are the state and action spaces respectively, 
    \item $P:\mathcal{S}\times\mathcal{A}\times\mathcal{S}\rightarrow[0,1]$ is the transition function capturing the environment dynamics,
    \item $r:\mathcal{S}\times\mathcal{A} \rightarrow \mathbb{R}$ is the reward function associated with each state-action pair,
    \item $\gamma$ is a discount factor in the range $[0,1)$, 
    \item $\mathcal{S}_I\subseteq \mathcal{S}$ is the initial state set,
    \item $\Delta_0:\mathcal{S}_I\rightarrow (0,1]$ is the initial state distribution, and
    \item $\pi:\mathcal{S}\times\mathcal{A}\rightarrow [0,1]$ is a stochastic policy that is a distribution capturing an action distribution given a state. Actions are sampled from this policy and affect the environment defined by the MDP.
\end{itemize}
In unconstrained reinforcement learning, the goal is to learn an optimal policy $\pi^*$ maximizing expected discounted sum of rewards along a trajectory:
\begin{align}
        \pi^*=\arg\max_\pi \E_{s\sim \Delta_0} V^\pi_r (s), \mbox{where}\\
    V^\pi_r(s):=\E_{\xi \sim \pi,P(s)} \biggl[\sum_{s_t\in \xi} \gamma^{t} r(s_t,a_t)\biggr].
\end{align}
Here, $\xi \sim \pi,P(s)$ indicates sampling trajectory $\xi$ for horizon $T$ starting from state $s$ using policy $\pi$ in the MDP with transition model $P$, and $s_t\in \xi$ is the $t^{th}$ state in trajectory $\xi$. Similarly, $s' \sim \pi,P(s)$ indicates sampling the next state after state $s$ using policy $\pi$ with transition model $P$. We will use the notation $s'$ to mean by default the next (sampled) state after the state $s$.

\subsection{Dynamical Systems and HJ Reachability}
In this paper, we will consider continuous, fully observable dynamics that are either deterministic or stochastic with bounds. Consider a dynamical system $f:\mathcal{S}\times\mathcal{A}\times\mathcal{D}\rightarrow{S}$:
\begin{equation}
    \frac{ds}{dt} = f(s,a,d)
\end{equation}
in which the state is $s\in \mathcal{S}\subseteq \mathbb{R}^n$, the control (also known as action) is $a\in \mathcal{A}$, and the disturbance is $d\in \mathcal{D}$, where $\mathcal{A}\subseteq \mathbb{R}^{m_a}$ and $\mathcal{D}\subseteq\mathbb{R}^{m_d}$ are compact sets. We assume $f$ is Lipschitz continuous in $s$ and uniformly bounded. We also assume that the control and disturbance signals $a(\cdot)$ and $d(\cdot)$ are measurable (for a precise definition of measurable see Chapter 17 of~\citet{carothers2000real}). In most cases, the works we cover either do not have a disturbance variable, or model disturbance as a random sampled value. If there is no disturbance, then the dynamical model is simply $f:\mathcal{S}\times\mathcal{A}\rightarrow{S}$.

Consider a Lipschitz surface function $h:\mathcal{S}\rightarrow \mathbb{R}^{\geq 0}$ which is the safety loss function that maps a state to a non-negative real value, which is called the constraint value, or simply cost. Note that $h(s)=0$ if and only if there is no constraint violation at state $s$.

The failure set $\mathcal{F}$ is the set of states for which there is an instantaneous constraint violation. Formally, the failure set is defined as the super-zero level set of $h$. In particular,
\begin{equation}
    s\in\mathcal{F} \iff h(s)> 0.
\end{equation}
On the other hand, a target set is the set of states for which it is desirable to reach, and it can be similarly defined. We will explore target sets in more depth in reach-avoid problems in Section~\ref{sec:ReachAvoid}.

For a deterministic dynamics, it is possible to determine if an initial state will lead to failure despite optimal actions. Then, the value function $V:\mathcal{S} \times \mathbb{R} \rightarrow \mathbb{R}$ and associated reachable set $\mathcal{R}(\mathcal{F}, t)$ are defined as:
\begin{align}
\label{eq:OptValDef}
    V(s,t) &:= \sup_{d(\cdot)} \inf_{a(\cdot)} \sup_{\tau\in [t,T]} h(s_{\tau})\\
    \mathcal{R}(\mathcal{F}, t) &:= \{s\in \mathcal{S} : V(s, t) > 0\}. \label{eq:avoidset}
\end{align}

In effect, this optimization over the action signal minimizes the maximum possible reachable violation starting from any point in the state space. If the control never enters the failure set when starting from state $s$, the value function will be zero. Otherwise, the value function will be strictly positive. In the case of a finite horizon in time interval $t\in [0,T]$, dynamic programming can obtain the optimal control and value function. Specifically, this will be the solution to the time-dependent terminal-value Hamilton-Jacobi-Bellman variational inequality (HJBVI)~\cite{akametalu2018minimum}:
\begin{multline}
\label{eq:hjbvi}
0=\max \bigg\{ h(s) - V(s,t), \frac{\partial V}{\partial t} + \min_{a\in \mathcal{A}} \max_{d\in \mathcal{D}} \nabla_{s}V^{\top} f(s,a,d)   \bigg\}, \\
V(s,T)=h(s), \forall s\in \mathcal{S}.
\end{multline}

Now as $T\rightarrow\infty$, if $V$ converges to a fixed solution then $V(s,t)$ will be independent of $t$. Thus the time parameter can be dropped to obtain the optimal value function $V(s)$.

\section{Traditional HJ Reachability for Learned Controls}
\label{sec:TradHJ_RL}

We first briefly discuss traditional HJ reachability analysis techniques for reinforcement learning-based control. Recent papers propose approaches that evaluate the safety (or probe the safe space) of learning-based control by analytically computing solutions of the dynamics's HJBVI. These methods require having access to or reconstructing the system's model dynamics. With a model, approaches can compute gradients of the dynamics at any given state.

The work of~\citet{akametalu2014reachability} makes inferences about disturbances to perform reachability analysis. Particularly, the work uses Gaussian processes to construct the disturbance set from previous observations of the dynamics and then solve the HJBPDE to compute an optimally safe control and safety value function. Then, a safe framework can be defined using any safety-aware learned (task-solving) control and this optimally safe control and safety value function. Namely, whenever the value function satisfies some safety threshold, then the safety-aware learned control is deployed. Otherwise, the default optimally safe controller is used.

Another work~\cite{fisac2018general} employs model-based HJ reachability analysis in conjunction with Bayesian-inference techniques to create a safety framework that can incorporate an arbitrary learning-based control algorithm. When there are no safety concerns, it permits a learned control policy to optimize for a particular task. Else it defaults to a safe policy computed via solving the HJBPDE. The safety choice of picking between these two policies is determined via safety analysis refined through Bayesian inferences from online data, particularly using Gaussian processes.

\citet{ivanovic2019barc} is a model-based approach based on backward reachability. In particular, it iteratively uses backward reachability (also known as inverse problem in the theoretical literature~\cite{colombo2020initial, esteve2020inverse}) from the final goal states to construct a set of initial state distributions under some approximate deterministic model dynamics with no disturbance consideration. Then, at each iteration, it proposes using typical model-free reinforcement learning methods to acquire a policy to get from an initial state (sampled uniformly from a growing backward reachable set) to the goal under a potentially stochastic dynamics.

In the rest of this survey, we will primarily discuss learning-based methods for obtaining the HJ reachability value function via reinforcement learning. We term this technique as HJ reachability estimation.

\section{Learning Reachability in Model-free Settings}
\label{sec:LearnHJ}
Overcoming the computational complexity of traditional HJ reachability analysis methods requires a scalable approach to acquire the HJ reachability value function. The recent literature has proposed a new direction of approximating the HJ reachability value function through learning-based approaches in the face of unknown dynamics. In particular, similar to a reward or cost critic, an HJ reachability function can be learned in an online, recursive fashion. Within the reinforcement learning framework, we can construct algorithms that obtain reachable sets via a data-driven, sampling-based manner that is 1) generalizable, since there is no need for direct access to the dynamics, and 2) scalable, in part due to the guaranteed convergence to a unique value function solution with gamma contraction mapping.

\subsection{Bellman Formulation}
\label{subsec:Bellman_form}

To learn an estimation of the HJ reachability value function in an online fashion, the value function must be equivalently defined with a backup operator in the form of the recursive Bellman update.

In particular, the works of~\citet{akametalu2018minimum, fisac2019bridging} demonstrate that the discrete approximation solution of~\eqref{eq:hjbvi} with no disturbances is:
\begin{equation}
    V(s,t) = \max \bigg\{ h(s), \min_{a\in \mathcal{A}} V(s + f(s,a)\Delta t, t + \Delta t) \bigg\}.
\end{equation}

Furthermore, as $T\rightarrow \infty$, if $V$ converges, then $V$ does not change with respect to time, so it satisfies the Bellman equation:
\begin{align}
    V(s) &= \max \{ h(s), \min_{a\in \mathcal{A}} V(s + f(s,a)\Delta t) \} \\
         &= \max \{ h(s), \min_{a\in \mathcal{A}} V(s') \}
\end{align}
where $s'$ is the next state after $s$ in the trajectory. Using this Bellman reformulation, the HJ reachability value function of the optimal control can be learned using the recursive dynamic programming approach known as value iteration. Notice that if this method is used to obtain a value function and optimal policy in a stochastic setting (i.e. the transition function and/or the policy is probabilistic) it would return a value function capturing the expected maximum cost along a trajectory sampled from the policy and transition function. This value function is not useful or well-defined for hard constraint tasks since a stochastic policy will likely enter a violation with some non-zero probability when starting from most states.

Nonetheless, it is still possible to use the Bellman recursive formulation for acquiring the HJ reachability value function to learn a meaningful tool for stochastic MDPs and policies using a special cost function~\cite{abate2008probabilistic, ganai2023iterative}. Consider the binary indicator cost function $\mathbbm{1}_{h(s)>0}$ which returns $1$ if there is a constraint violation at state $s$, and returns $0$ otherwise. In this setting, the optimal control $\pi: \mathcal{S}\times\mathcal{A}\rightarrow [0,1]$ is the one that minimizes the likelihood of entering the set of constraint violation states along the trajectory under the stochastic MDP with transition likelihood function $P$. Formally, in the discrete-time setting, the optimal control and its associated value function $\phi:\mathcal{S}\rightarrow [0,1]$, called the reachability estimation function (REF), are defined by~\cite{ganai2023iterative, abate2008probabilistic}:
\begin{equation}
    \phi(s) := \inf_{\pi(\cdot | \cdot)} \E_{\xi\sim \pi, P(s)} \sup_{s_t \in \xi} \mathbbm{1}_{h(s)>0}.
\end{equation}

Although the value function is defined for stochastic dynamics (notice the expectation over the sampled trajectories),~\citet{ganai2023iterative} exploits the binary nature of the instantaneous cost indicator function to create a Bellman recursive formulation of the REF:
\begin{equation}
    \phi(s) = \max \bigg\{ \mathbbm{1}_{h(s)>0},  \min_{\pi(\cdot | s)} \E_{s'\sim \pi, P(s)} \phi(s') \bigg\}.
\end{equation}
When this value function is learned for a particular control it can provide information on the probability that the control at any given state will reach a violation. 

\subsection{Discounted HJ value function for Reinforcement Learning}
\label{subsec:DiscRLHJ}

Temporal difference learning is a preeminent class of model-free reinforcement learning algorithms that estimates the value function for a particular control policy. In other words, the value function $V^{\pi}(s)$ with Bellman operator $\mathcal{B}^{\pi}$ (i.e. the operator that defines the recursive Bellman formation), should be estimated for a particular control policy $\pi$.
This can be done by iteratively updating the value function with the temporal difference rule using trajectory samples collected online. At update $k$, for learning rate $\alpha$, the temporal difference rule is~\cite{SuttonRL, sutton1988learning, tsitsiklis1994asynchronous}:
\begin{equation}
    V_{k+1}^{\pi}(s) \leftarrow V_{k}^{\pi}(s) + \alpha(\mathcal{B}^{\pi} V_{k}^{\pi}(s) - V_{k}^{\pi}(s)).
\end{equation}

In order to guarantee convergence to the unique solution of the Bellman equation, the Bellman operator $\mathcal{B}^{\pi}$ must induce a gamma contraction mapping in the space of value functions~\cite{denardo1967contraction}. In general, time-discounting in the Bellman formulation of the value function enables the reachable set to be estimated as a fixed point in a contraction mapping~\cite{akametalu2018minimum}.

To address this, the approach found in~\citet{akametalu2018minimum} proposes a modified discounted optimal control value function. For the defined cost function $h:\mathcal{S}\rightarrow \mathbb{R}^{\geq 0}$, the optimal control and value function are defined by:
\begin{equation}
\label{eq:disc_optform}
    V(s) := \inf_{\pi(\cdot)} \sup_{t\geq 0} h(s_t) e^{-\lambda t}
\end{equation}
for some discount rate $\lambda\in\mathbb{R}^{>0}$.

Similar to the non-discounted Bellman formulation, this value function and its optimal control can be obtained by solving the Hamilton-Jacobi-Bellman variational inequality~\cite{akametalu2018minimum}:
\begin{equation}
    \label{eq:hbjvidiscounted}
    0=\max \bigg\{ h(s) - V(s,t), \min_{a\in \mathcal{A}} \nabla_{s}V^{\top} f(s,a) - \lambda V(x)   \bigg\}.
\end{equation}
This has the discrete-time solution:
\begin{align}
\label{eq:hjdiscsol}
    V(s) = \max \{ h(s), \min_{a\in \mathcal{A}} \gamma V(s') \}
\end{align}
where $\gamma=e^{-\lambda \Delta t}$ is the discount factor. The authors demonstrate the gamma contraction mapping for this discounted Bellman formulation for $\gamma\in(0,1)$, and thereby guarantee that temporal difference learning will converge to the unique value function solution.

The work of~\citet{fisac2019bridging} proposes a different Bellman formulation for learning an estimation of the HJ reachability value function:
\begin{align}
\label{eq:hjdiscsol_bridging}
    V(s) = (1-\gamma)h(s)+\gamma\max \{ h(s), \min_{a\in \mathcal{A}}  V(s') \}.
\end{align}
While this value function is not an exact discrete-time solution of the HJBVI in~\eqref{eq:hbjvidiscounted}, the work of~\citet{fisac2019bridging} proves this provides a tighter gamma contraction mapping than~\eqref{eq:hjdiscsol}, and therefore temporal difference learning can converge to the value function solution faster. Notice that using the cost function as the binary indicator function $\mathbbm{1}_{h(s)>0}$ in lieu of $h(s)$ would make~\eqref{eq:hjdiscsol} and~\eqref{eq:hjdiscsol_bridging} become identical Bellman formulations.

Using the discounted Bellman formulations, HJ reachability can be incorporated into reinforcement learning problems. In~\citet{fisac2019bridging}, the authors use the HJ reachability value function as the critic and the policy optimization algorithm REINFORCE~\cite{williams1992simple} to solve control problems in environments like the lunar lander and the $18$-dimensional jumping half-cheetah.

\section{Solving Reach-Avoid Problems}
\label{sec:ReachAvoid}
Reach-avoid problems form a class of environments in which the goal is to control the agent to reach a target set of states while simultaneously avoiding a failure set of states~\cite{fisac2015reach, barron1990differential, mitchell2005time, margellos2011hamilton, bokanowski2011minimal}. We have previously discussed how HJ reachability has been used to solve the avoidance problem. Recent literature has demonstrated how to combine the reach problem and the avoid problem in HJ reachability simultaneously, as well as how to combine HJ reachability with other control theoretic functions to solve the reach-avoid problem in the online setting.

\subsection{Learning HJ Reach-Avoid Value Function}
The work of~\citet{fisac2015reach} establishes how to formally define reach-avoid problems. Specifically, the problem seeks to find the optimal control such that given a starting state, the agent can reach the target set of states $\mathcal{T}$ while avoiding the failure set of states $\mathcal{F}$. They define two cost functions $l:\mathcal{S}\rightarrow \mathbb{R}$ and $g:\mathcal{S}\rightarrow \mathbb{R}$ such that for any state $s\in S$:
\begin{align}
\label{eq:lg_prop}
\begin{split}
    l(s)\leq 0 &\iff s\in \mathcal{T} \\
    g(s) > 0 &\iff s\in \mathcal{F}.    
\end{split}
\end{align}

Then with deterministic MDP, in discrete time, for a finite horizon time $T$, a payoff function for a deterministic control policy $\pi:\mathcal{S}\rightarrow\mathcal{A}$ can be defined as:
\begin{equation}
\label{eq:payoff}
    \mathcal{V}^{\pi}(s, T)=\min_{t\in[0...T]} \max \bigg\{ l(s_t), \max_{\tau\in[0...t]} g(s_\tau)  \bigg\}.
\end{equation}
The outer maximum considers the possibility of ever reaching the target set. The inner maximum ensures that, during the time taken to reach the target set, there are no states in the trajectory that are in the failure set. Thus, for a given time $T$, if there exists a time $t$ when the agent reaches a state $s_t$ in the target set while avoiding the failure set, then the payoff function will be at most $l(s_t)\leq 0$ and therefore non-positive. However, if the agent always enters the failure set before the target set, then at any time $t$, there would always exist a time $w\in[0...t]$ such that $g(s_w)>0$, and therefore the payoff is positive. Step-wise noise disturbance can be considered within the payoff function, and a dynamic programming value iteration approach to obtaining the payoff function for a particular control can be formulated~\cite{fisac2015reach}.

Consider infinite horizon (i.e. $T\rightarrow\infty$). For the sake of simplifying notation, we can define: 
\begin{equation}
    \mathcal{V}^{\pi}(s)=\lim_{T\rightarrow \infty} \mathcal{V}^{\pi}(s, T).
\end{equation}

As shown in a subsequent work~\cite{hsu2021safety}, the optimal control and its associated value function can then be defined as the one that minimizes the payoff function of~\eqref{eq:payoff}:
\begin{equation}
    V(s)=\inf_{\pi(\cdot)} \mathcal{V}^{\pi}(s).
\end{equation}

Observe that the sign of the payoff function can tell us if the control signal starting from state $s$ will satisfy the reach-avoid condition. So, if and only if $V(s)\leq0$, then there exists a control that can solve the reach avoid problem starting from state $s$.

Now, just as in the case for model-free learning of the HJ reachability function in Section~\ref{subsec:DiscRLHJ}, it is possible to learn the optimal HJ reach-avoid function.~\citet{hsu2021safety} provides a discounted (recall the importance of gamma contraction mapping) reach-avoid Bellman formulation suitable for learning online with temporal difference learning. Specifically, 
\begin{align}
\begin{split} \label{eq: discounted-reach-avoid}
    V(s) &= (1-\gamma)\max\{l(s),g(s)\} \\
    &+\gamma\max\bigg\{ \min \big\{l(s), \min_{a\in\mathcal{A}} V(s') \big\}, g(s)\bigg\}
\end{split}
\end{align}
where $s'$ is the next state produced by the MDP upon taking action $a$ from state $s$.

With this recursive reformulation of the value function,~\citet{hsu2021safety} uses the standard reinforcement learning algorithm Deep Q-Network (DQN)~\cite{mnih2013playing} to obtain the corresponding optimal control policy. They test this algorithm on environments such as an attack-defense game with two Dubins cars, and the Lunar Landing environment.

\subsection{Combing Reachability with Control Lyapunov for Stabilize-Avoid Problems}
Within the class of reach-avoid problems are the stabilize-avoid problems, in which the goal is to find a control that avoids the failure set while stabilizing toward the target set. If the target set consists of equilibrium points, then standard reach-avoid algorithms can be used to solve the stabilize-avoid problems. However, in many cases, the target set may additionally consist of non-equilibrium points. To use the reach-avoid algorithms in the stabilize-avoid problem in this general case, the set of equilibrium points must be extracted from the target set. This extraction is difficult and may even be impossible if such a set does not exist. HJ reachability-inspired approaches can be combined with the control Lyapunov function to solve Stabilize-Avoid problems. 

In the work of~\citet{so2023solving}, the stabilize-avoid problem is formulated as a constraint optimization problem. Particularly, for a deterministic MDP and using the cost functions $l:\mathcal{S}\rightarrow\mathbb{R}^{\geq 0}$ and $g:\mathcal{S}\rightarrow\mathbb{R}$ with properties of~\eqref{eq:lg_prop}, the undiscounted value function for policy $\pi$ is defined along the trajectory as:
\begin{equation}
    V^{l,\pi}(s) := \sum_{t=0}^{\infty} l(s_t)
\end{equation}
where $\{s_t\},t\in \mathbb{Z}^{\geq 0}$ is the trajectory under $\pi$ starting from state $s=s_0$. Furthermore, the optimal control problem is defined as:
\begin{align}
\begin{split}
\min_{\pi}& V^{l,\pi}(s) \\
\text{s.t. }& g(s_t) \leq 0, \forall t\geq0.
\end{split}    
\end{align}
Under some assumptions based on bounding the cost function $l$ and its dynamics under control $\pi$ by some state measure,~\citet{so2023solving} proves that $V^{l,\pi}$ is a Lyapunov function. They also convert the constraint problem into the epigraph form~\cite{boyd2004convex}:
\begin{align}
\label{eq:epigraph}
\begin{split}
&\min_{z} z \\
\text{s.t. } 0\geq \min_{\pi} \max\bigg\{ &\max_{t\in\mathbb{Z}^{\geq0}} g(s_t), V^{l,\pi}(s)-z \bigg\}.
\end{split}    
\end{align}
In effect, $z$ acts as the accumulated $l$ cost budget, and the goal is to minimize the maximum needed cost budget and ensure the agent avoids entering the failure set where $g(s)>0$. The RHS of the constraint in this epigraph form can be learned as a value function parameterized by both the state and the cost budget. Namely,~\citet{so2023solving} learns this optimal control value function by applying a recursion similar to~\eqref{eq: discounted-reach-avoid}:
\begin{equation}
    V(s,z)=\min_{a\in \mathcal{A}} \max\{g(s), V(s', z-l(s))\}.
\end{equation}
The algorithm uses a standard policy gradient approach to learn this value function online, and then in a subsequent stage solves the problem of~\eqref{eq:epigraph} by training via regression a neural network $z(s)$ that minimizes $V(s,z(s))$. This approach has been used to solve various complex stabilize-avoid problems including a $17$ dimension F16 fighter jet~\cite{heidlauf2018verification} ground collision avoidance in a low-altitude corridor.

\section{Model-free Safe Reinforcement Learning}
\label{sec:ModelFreeSafeRL}

Safe reinforcement learning is a setting in which the goal is to maximize some cumulative rewards while constraining the costs (i.e. constraint violations) along a trajectory~\cite{gu2022review, zhao2023state, garcia2015comprehensive, brunke2022safe}. In previous sections, the problems were reduced to optimizing a single (potentially composite) value function. However, in safe reinforcement learning, the problem generally requires keeping track of two separate value functions, one for rewards and another for costs, and optimizing a composite expression involving both value functions. The reward value function $V^{\pi}_r$ is specifically defined as the discounted cumulative rewards found in Section~\ref{subsec:MDP}. However, the cost value function's definition is determined by the specific optimization framework. 

Traditionally, safe reinforcement learning was solved within the constrained Markov decision process (CMDP) framework~\cite{CMDP} in which the cost value function was the discounted cumulative costs similar to the reward value function:
\begin{equation}
\label{eq:cum_cost}
    V^\pi_c(s):=\E_{\xi \sim \pi,P(s)} [\sum_{s_t \in \xi} \gamma^t h(s_t)].
\end{equation}
Then, for some environment-defined positive cost threshold $\chi$, the CMDP-constrained optimization takes the form:
\begin{align}
\tag{CMDP}
\begin{split}
\label{eq:CMDP}
    \max_\pi& \E_{s\sim \Delta_0}[V^\pi(s)] \\
    \text{s.t. }&  \E_{s\sim \Delta_0}[V^\pi_c(s)] \leq \chi.
\end{split}
\end{align}

Various approaches have been proposed to solve safe reinforcement learning in this framework. Trust-region approaches~\cite{YangProj, zhang2020first, yang2022cup, achiam2017constrained} try to guarantee monotonic improvement in performance while ensuring constraint satisfaction. Primal-dual approaches~\cite{tessler, ma2021feasible, ray2019benchmarking, duan2022adaptive} use Lagrangian relaxation of the constraints to optimize an expression involving the reward and cost value functions. Outside of these two classes exist approaches like constraint-rectified policy optimization (CRPO)~\cite{xu2021crpo}, which takes a policy gradient update step toward improving $V^{\pi}_r$ if constraints are satisfied at a particular iteration, otherwise it takes steps to minimize $V^{\pi}_c$. This approach guarantees convergence to optimum under certain assumptions.

The main drawback of the CMDP framework is its lack of rigorous guarantees of persistent safety. Because the framework permits some positive amount of constraint violations ($\chi>0$), it cannot be used for state-wise constraint optimization problems. Another issue is that choosing a cost threshold $\chi$ for an environment requires tuning and/or prior familiarity with the environment. To address this, recent literature has proposed methods of using the safety guarantees provided by Hamilton-Jacobi reachability to redefine the problem into a constrained optimization within feasible (i.e. constraint-satisfying) states. We explore recent algorithms with frameworks for the deterministic and stochastic dynamics cases.

\subsection{Deterministic Safe Reinforcement Learning}

When the MDP is deterministic, the HJ reachability value function can be learned online through the Bellman update from~\eqref{eq:hjdiscsol_bridging}. Specifically, for a control policy $\pi$, define the HJ reachability value function recursively as:
\begin{equation}
    V^{\pi}_h(s)=(1-\gamma)h(s)+\gamma\max\{h(s), V^{\pi}_h(s')\}.
\end{equation}

The reachability value function is used to probe whether a state is within the \textit{feasible} set, which is the set of states starting from which the agent will never enter the failure (i.e. constraint violating) set(s) along its trajectory. Formally, for a particular control $\pi$, and its associated reachability value function $V^{\pi}_h$, the feasible set is defined as:
\begin{equation}
    \mathcal{S}^{\pi}_f:=\{s\in \mathcal{S} : V^{\pi}_h(s)=0 \}.
\end{equation}

Some papers refer to this feasible set as the safe set, and is the complement of  $\mathcal{R(F)}$ from \eqref{eq:avoidset}. By learning the reward value function $V^{\pi}_r$ and reachability value function $V^{\pi}_h$, a recent approach~\cite{chen2021safe} solves safe control tasks by considering the two cases of whether a state is feasible or not and learning a different control for each case. Similar to the CRPO algorithm, during training, if the state is in the feasible set (with some tolerance $\epsilon$) then an action is taken from the control that optimizes $V^{\pi}_r$ and that control is updated. Otherwise if the state is infeasible, then an action is taken from the "safe" control which minimizes the maximum reachable violation, i.e. $V^{\pi}_h$, and this safe control is updated. This technique falls within the broader class of shielding~\cite{fisac2018general}, which is discussed in more detail in Section~\ref{sec:RobustRW}. This approach is notable for solving a high-dimensional, vision-based autonomous racing environment called Learn-to-Race~\cite{herman2021learn}. 

However, to fully address the problems of CMDP (lack of safety guarantees stemming from tolerance of some constraint violation), environment-specific cost thresholds/tolerance should be avoided altogether. Instead, the recent literature~\cite{yu2022reachability,ganai2023iterative} has moved toward learning optimal (largest) feasible sets. The largest feasible set can be defined as:
\begin{equation}
    \mathcal{S}_f:=\{s\in \mathcal{S} : \exists \pi, V^{\pi}_h(s)=0 \}.
\end{equation}
In other words, the largest feasible is the set of states for which there exists a control policy that ensures no constraint violations along a trajectory starting from those states. The largest feasible set can also be written as:
\begin{equation}
    \mathcal{S}_f = \bigcup_{\pi} S^{\pi}_f.
\end{equation}
By obtaining or having access to this largest feasible set, the hope is that the algorithms can learn controls that overcome the conservative behavior seen in other control/energy-based approaches like CBFs~\cite{ma2021model, mayne2000constrained}.

Let the binary function $\mathbbm{1}_{s\in \mathcal{S}_f}$ indicate whether a state is in this largest feasible (returning $1$) or not (returning $0$). Then, the work of~\citet{yu2022reachability} proposes a novel optimization framework that considers optimization under two scenarios depending on whether the state is in $\mathcal{S}_f$, assuming one has access to this oracle $\mathbbm{1}_{s\in \mathcal{S}_f}$. In particular, if state $s\in \mathcal{S}_f$, the goal would be to optimize for maximum reward value function starting from that state under the constraint that the trajectory continues to persistently remain within the feasible set (and thereby incur no future violations). On the other hand, if the state $s\notin \mathcal{S}_f$, then the goal is to find a control that minimizes the maximum reachable violation starting from that state. Formally, this optimization called Reachability Constrained Reinforcement Learning (RCRL) can be expressed as:
\begin{align}
\tag{RCRL}
\label{eqn:RCRL}
\begin{split}
    \max_\pi &\E_{s\sim \Delta_0} [V^{\pi}_r(s)\cdot \mathbbm{1}_{s\in \mathcal{S}_f} - V^{\pi}_h(s)\cdot \mathbbm{1}_{s\notin \mathcal{S}_f}]\\
    \text{s.t. } & V^{\pi}_h (s)\leq 0, \forall s\in \mathcal{S}_I \cap \mathcal{S}_f.
\end{split}
\end{align}
The Lagrangian of~\eqref{eqn:RCRL} can be formulated as:
\begin{align}
\label{eq:RCRLLag}
\begin{split}
    \mathcal{L}(\pi, \lambda) = \E_{s\sim \Delta_0} [V^{\pi}_r(s)\cdot \mathbbm{1}_{s\in \mathcal{S}_f} - V^{\pi}_h(s)\cdot \mathbbm{1}_{s\notin \mathcal{S}_f}] \\
    +\int_{\mathcal{S}_f \cap \mathcal{S}_I} \lambda(s) V^{\pi}_h(s)ds.
\end{split}    
\end{align}

The main challenge in solving this optimization is being able to acquire the \textit{largest} feasible set. To overcome this,~\citet{yu2022reachability} solves their optimization by providing guarantees in stochastic gradient descent optimization of the policies, critics, and Lagrangian multiplier via the stochastic approximation theory framework established in~\citet{borkar2009stochastic, chow2017risk}, and used in~\citet{chow2019lyapunov}. 

~\cite{yu2022reachability} proposes finding a saddle point of the surrogate Lagrangian optimization of~\eqref{eqn:RCRL} as:
\begin{equation}
\label{eq:RCRL_Lag_sur}
    \min_{\pi}\max_{\lambda} \E_{s\sim \Delta_0} [-V^{\pi}_r(s) + \lambda(s)V^{\pi}_h(s)].
\end{equation}
The idea behind this formulation is that $\lambda(s)$ will eventually converge to a finite value for feasible states and diverge for infeasible states~\cite{ma2021feasible}. Recall that for feasible states $s$, $V^{\pi}_h(s)=0$, so the optimization becomes simply minimizing $-V^{\pi}_r(s)$ regardless of the magnitude of $\lambda(s)$. However, for infeasible states, $V^{\pi}_h(s)>0$, so the optimization minimizes $-V^{\pi}_r(s)+\lambda V^{\pi}_h(s)$ for very large $\lambda$. Notice, however, that since the Lagrangian multiplier diverges for infeasible states, $-V^{\pi}_r(s)$ can be ignored. So, the optimization is effectively minimizing $V^{\pi}_h(s)$.

If $\lambda(s)$ is the Lagrangian multiplier for the optimal control, then solving the surrogate Lagrangian optimization in~\eqref{eq:RCRL_Lag_sur} is equivalent to solving the Lagrangian of~\eqref{eq:RCRLLag}.~\citet{yu2022reachability} demonstrates this can be achieved primarily by configuring the learning rate schedules of the learned networks. Say, the critics maintain a step size schedule of $\{\zeta_1(k)\}$, the policy maintains a step size schedule of $\{\zeta_2(k)\}$, and the Lagrangian multiplier maintains a step size schedule of $\{\zeta_3{k}\}$ for iteration $k$. Based on stochastic approximation theory~\cite{borkar2009stochastic, chow2017risk}, if:
\begin{align}
\label{eq:RCRL_lrassump}
\begin{split}
    \sum_k \zeta_i(k)=\infty \text{ and } \sum_k \zeta_i(k)^2<\infty, \forall i\in \{1,2,3\} \\
    \text{ and } \zeta_3(k) = o(\zeta_2(k)), \zeta_2(k)=o(\zeta_1(k)),
\end{split}
\end{align}
then it is possible to prove that the updates of the critic, policy, and Lagrangian multiplier will result in convergence of the local optimal policy of~\eqref{eqn:RCRL} \textit{almost surely} (i.e. with likelihood $1$). The reward and cost critic networks have a faster learning rate schedule than the policy networks and therefore converge to the current policy's optimal value functions. The Lagrangian multiplier network has a learning schedule slower than the policy network and therefore can be thought of as capturing the overall trends of feasibility. If during training there was a policy that was able to make a particular state in its feasible set, then $\lambda(s)$ will capture that information. If in the future, the policy no longer makes the state in the feasible set, the Lagrangian multiplier will increase and thereby penalize the policy. Using this approach,~\citet{yu2022reachability} is able to solve hard constraint problems in the Safety Gym~\cite{ray2019benchmarking} environment with static hazards and obstacles.

\subsection{Stochastic Safe Reinforcement Learning}

Under a stochastic MDP, HJ reachability can still be a useful tool for guaranteeing optimal control with safety guarantees. We present in Section~\ref{subsec:Bellman_form} how recent works define a HJ reachability value function called the Reachability Estimation Function (REF) for a binary cost function $\mathbbm{1}_{h(s)>0}$ under stochastic dynamics. The optimal REF captures the minimum likelihood of entering the set of constraint violation states. In effect, the REF is the likelihood that a state is \textit{infeasible} -- we will thus use the phrase \textit{likelihood of feasibility} to mean $1-\phi(s)$ and \textit{the likelihood of infeasibility} to mean $\phi(s)$.

The work of~\citet{ganai2023iterative} proposes to use the REF function in defining the optimization formulation. In particular, in place of the deterministic feasibility indicator $\mathbbm{1}_{s\in \mathcal{S}_f}$ they use the likelihood of feasibility $1-\phi(s)$, and instead of the deterministic infeasibility indicator $\mathbbm{1}_{s\notin \mathcal{S}_f}$ they use the likelihood of infeasibility $\phi(s)$. Note these feasibility sets are the largest/optimal.

However, simply replacing the indicator function with $\phi(s)$ in the optimization of~\eqref{eqn:RCRL} will not be a valid construction for the stochastic case since $V^{\pi}_h$ is not well defined for stochastic dynamics.~\citet{ganai2023iterative} addresses this by using the cumulative cost function $V^{\pi}_c$ as defined in the CMDP framework in~\eqref{eq:cum_cost}. In particular, they replace $V^{\pi}_h$ with $V^{\pi}_c$ in~\eqref{eqn:RCRL}.

In the constraint, $V^{\pi}_c(s)\leq 0$ is satisfied if and only if persistent safety (i.e. no constraint violations along the trajectory) is guaranteed for that state under control policy $\pi$. Therefore, $V^{\pi}_c(s)\leq 0$ can be used as a valid measure for constraining the agent to remain within the feasible set.

Furthermore, $V^{\pi}_c$ provides important safety guarantees when the agent is in the infeasible set. Specifically,~\citet{ganai2023iterative} proves that an optimal control minimizing $V^{\pi}_c$ can verifiably \textit{enter} the feasible set when starting in the infeasible set if there exists a control given sufficient time. Intuitively, consider that $V^{\pi}_c(s)$ is the (average) cumulative cost of a trajectory starting at $s$ (ignore the discount factor by making say $\gamma=1$). If the control enters the feasible set, $V^{\pi}_c(s)$ is finite since there will be a point after will no more costs are accumulated. Otherwise if the control remains in the infeasible set, then $V^{\pi}_c(s)$ is infinite since there will always be costs accumulated at some points in the trajectory. Thus, if there exists a control that enters the feasible set at state $s$, then the minimum cumulative cost for a policy starting from state $s$ is finite, and thus the optimal control minimizing $V^{\pi}_c(s)$ will enter the feasible set.~\citet{ganai2023iterative} provides a proof along these lines with consideration to the discount factor $\gamma\in [0,1)$.

Using the REF and the cumulative cost value function,~\citet{ganai2023iterative} proposes an optimization formulation for safety constraint reinforcement learning that works for both stochastic and deterministic environments. Formally, their optimization called Reachability Estimation for Safe Policy Optimization (RESPO) can be expressed as:
\begin{align}
\tag{RESPO}
\label{eq:RESPO}
\begin{split}
    \max_\pi \E_{s\sim \Delta_0} [V^{\pi}_r(s)\cdot (1-\phi(s)) - V^{\pi}_c(s)\cdot \phi(s)] \\
    \text{ s.t. }V^{\pi}_c (s) \leq 0, \text{ w.p. } 1-\phi(s), \forall s\in S_I.
\end{split}
\end{align}

To learn the value function online, they create a discounted Bellman formulation to ensure gamma contraction mapping to demonstrate convergence to the solution (Section~\ref{subsec:DiscRLHJ}). Thus, they define a discounted Bellman formulation of the REF as:
\begin{equation}
    \phi(s)=\max\{\mathbbm{1}_{h(s)>0}, \gamma\min_{a\in\mathcal{A}}\E_{s'\sim P(s,a)} \phi(s') \}.    
\end{equation}

The Lagrangian of~\eqref{eq:RESPO} is formulated as:
\begin{align}
\begin{split}
    \E_{s\sim \Delta_0}\biggl[ [-V^{\pi}_r(s) + \lambda\cdot V^{\pi}_c(s)]\cdot (1-\phi(s)) + V^{\pi}_c(s)\cdot \phi(s)\biggr].
\end{split}
\end{align}

Similar to the deterministic safe reinforcement learning approach~\eqref{eqn:RCRL}, the main challenge in solving the stochastic safe reinforcement learning approach~\eqref{eq:RESPO} is obtaining the optimal likelihood of entering the set of constraint violation states (i.e. the REF).~\citet{ganai2023iterative} proposes solving this problem via the stochastic approximation theory framework~\cite{borkar2009stochastic, chow2017risk}. Similar to~\eqref{eq:RCRL_lrassump}, say the learning rates of the critic value functions, the policy, REF, and lagrangian multiplier are $\{\zeta_1(k)\}$, $\{\zeta_2(k)\}$, $\{\zeta_3(k)\}$, and $\{\zeta_4(k)\}$ respectively. If we ensure:
\begin{align}
\label{eq:RESPO_lrassump}
\begin{split}
    \sum_k \zeta_i(k)=\infty \text{ and } \sum_k \zeta_i(k)^2<\infty, \forall i\in \{1,2,3,4\} \\
    \text{ and } \zeta_{i}(k) = o(\zeta_{i-1}(k)),\forall i\in \{2,3,4\},
\end{split}
\end{align}
then~\cite{ganai2023iterative} guarantees that the updates of the various learnable parameters will result in the policy network converging to the local optimal policy of~\eqref{eq:RESPO} \textit{almost surely} (it is important to note that the REF, learned like a value function, is on a \textit{slower} learning rate schedule than the policy!). The reasoning for guaranteed convergence is mostly similar to that of the deterministic safe reinforcement learning version~\eqref{eqn:RCRL}~\cite{yu2022reachability} except for the stochastic nature of the dynamics and $\phi$. In particular, since the learning rate schedule for the REF $\phi$ is slower than that of the policy,~\citet{ganai2023iterative} guarantees that $\phi$ will be the REF of the most optimal policy to the extent that the lagrangian multiplier $\lambda$ allows (since $\lambda$ is technically finite).~\eqref{eq:RESPO} learns stochastic policies that solve safety constrained problems in the Safe PyBullet framework~\cite{gronauer2022bullet}, MuJoCo~\cite{MuJoCo}, and Safety Gym~\cite{ray2019benchmarking} in which there are various moving/movable obstacles in addition to stationary hazardous regions. Furthermore,~\citet{ganai2023iterative} demonstrates how~\eqref{eq:RESPO} can incorporate and prioritize multiple hard and soft constraints to solve a multi-drone tunnel navigation environment.

\section{Robustness and Real-World settings}
\label{sec:RobustRW}

While most of the applications of Hamilton-Jacobi reachability we discussed so far solve problems in simulation, there has also been a line of work on learning verifiably safe controls in real-world settings. The main challenge in real-world settings is the presence of nondeterministic disturbances at each step. Take for instance quadrupedal robot control: the optimal control problem can be formulated as getting to region B in the fastest way possible, but other factors to consider include the presence of some unknown amount of wind or uncertain terrain. 

The recent literature solves this primary by constructing a safety filter~\cite{hsu2023safety} criterion $\Delta:\mathcal{S}\times\Pi\times\mathcal{Q}\rightarrow\{0,1\}$ dependent on the state $s\in\mathcal{S}$, the task solving (i.e. performance optimizing) control $\pi^{t}\in \Pi$, and backup optimally safe q-value function $Q^{u}\in \mathcal{Q}$. They can then define a composite policy $\pi^{sh}$ that uses the safety filter criterion $\Delta$ to decide whether to use the task-solving control $\pi^{t}$ or the backup optimally safe policy $\pi^u$ corresponding to $Q^{u}$. This approach of using the backup safe policy to override the tasking-solving policy is known as the least restrictive control law or shielding in~\citet{fisac2018general,alshiekh2018safe} and also examined in~\citet{cheng2019end, leung2020infusing}.

Hamilton-Jacobi reachability estimation methods have been used in constructing the safety filter criterion and/or the backup optimally safe policy. For instance, based on the work of~\citet{fisac2018general}, it is possible to construct the optimally safe q-value function in a Bellman formulation similar to that in~\eqref{eq:hbjvidiscounted}:
\begin{equation}
    Q^u(s,a) = (1-\gamma) h(s) +\gamma\max\big\{ h(s), \min_{a'\in \mathcal{A}} Q^u(s', a')  \big\}
\end{equation}
and define the safety filter criterion with an indicator function as:
\begin{equation}
    \Delta(s, \pi^{t}, Q^u):=\mathbbm{1}\big\{ Q^{u}(s,\pi^{t}(s)) \leq \epsilon \big\}
\end{equation}
for some threshold $\epsilon$. Then the composite policy can be formally constructed as:
\begin{align}
    \pi^{sh}(s) = \begin{cases}
  \pi^{t}(s), &  \Delta(s, \pi^{t},Q^u)=1\\
  \pi^{u}(s), & \text{otherwise.}
\end{cases}
\end{align}

\subsection{Fully Learning-based Control for Real-World Deployment}

Using this framework, it is possible to acquire policies that are (almost) ready to be deployed in real-world scenarios. One difficulty in deploying these algorithms is that learned control often struggles to generalize in new, unseen environments in the real world. To address this distributional shift between the simulation-based training data and the real-world testing data, the work of~\citet{hsuren2022slr} proposes a technique based on encouraging the generalization capabilities of the learned policies. They develop a $3$-tiered approach: learning control policies in Simulation, fine-tuning in a Lab, and then transferring the policies into the Real World. When training in Simulation, they use the HJ reachability-based shielding approach trained on RGB image vision-based observations. They augment this with a learning framework that optimizes for the diversity of robot learning behavior following the works of~\citet{eysenbach2018diversity, ren2021generalization}. The goal behavior in the simulation phase is to be able to reach the specified target through various paths. This can be done by conditioning the policy by some random latent variable representing a learned "skill" (i.e. taking a specific path to the target). By learning various ways (skills) to solve the problem, they can encourage the generalization capabilities of the learned control.

Subsequently, during the fine-tuning phase in the Lab environment, they can learn a prior distribution from which to sample the latent variables so as to find the best "skills," which were already learned in the simulation phase, needed to solve in some new lab environments.~\citet{hsuren2022slr} proposes doing this by leveraging the PAC-Bayes Control framework~\cite{majumdar2021pac, farid2022task, veer2021probably} to certify the generalization of the corresponding posterior distribution. Overall, this approach was tested on hardware experiments with the quadrupedal robot in real world indoor spaces. 

\subsection{Learning-based Control Shielded with Forward Reachability in Robust Deployment}

While learning-based control has the benefit of being scalable, the learned policy may not be accurate for all points in the state space and in general lacks intrinsic guarantees of safety. The work of~\citet{hsunguyen2023isaacs} addresses this problem by combining HJ reachability estimation and traditional HJ reachability analysis. Although they use a shielding framework similar to~\citet{fisac2018general, hsuren2022slr}, they learn a backup optimally safe controller that is disturbance aware and then define a new composite policy that includes the task solving policy $\pi^{t}$, the safe controller $\pi^u$, and an additional safe control policy based-on locally computing the forward reachability set.

To obtain the disturbance-aware backup controller, recent work considers the problem of obtaining a safe control policy that is resilient to the worst-case disturbance at each step. Specifically, while learning a control $\pi^{u}$ to solve the problem,~\citet{hsunguyen2023isaacs} proposes simultaneously treating the disturbance as an antagonist controlled with policy $\pi^{d}$. Then, in the typical game theoretic, adversarial fashion, the goal is to find a saddle point between both $\pi^u$ and $\pi^d$. Formally, the optimal controls and associated value function can be defined with the Bellman formulation:
\begin{equation}
    V(s) = (1-\gamma)h(s) + \gamma \min_{\pi^u} \max_{\pi^d} \E_{u,d} \max\big\{ h(s), V(s')  \big\}.
\end{equation}
The optimal control policies for this formulation are learned via the off-policy reinforcement learning algorithm Soft Actor-Critic algorithm~\cite{haarnoja2018soft}.

Even though these learned controls cannot provide intrinsic safety guarantees,~\citet{hsunguyen2023isaacs} constructs a composite policy that guarantees safety for $H$ horizon steps. In particular, they linearize the dynamics of the nominal local trajectory starting from state $s$ obtained from the learned control. Then, at some point $s'$ along the trajectory, they use a linear quadratic regulator approach to obtain a locally linear tracking policy $K(s'-s)$ for $H$ time into the future. Subsequently, they can define a safety criterion $\Delta:\mathcal{S}\times\Pi\times\mathbb{Z}^{\geq0}$. $\Delta(s,\pi^{t},H)=1$ if after applying one step of the task policy $\pi^{t}$, tracking policy $K$ can maintain safety under any disturbance for time horizon $H$ -- this safety is verified via forward HJ reachability analysis. Else $\Delta(s,\pi^{t},H)=0$. So, for a given state $s_t$ and future time step $\tau\in\{0...H\}$ along the nominal trajectory starting from $s_t$, the composite policy can be defined as:

\small
\begin{align}
  \pi^{sh}(s_{t+\tau}) = \begin{cases}
  \pi^{t}(s_t), &  \Delta(s_{t+\tau}, \pi^{t},H)=1\\
  K(s_{t+\tau}-s_t), & \Delta(s_{t+\tau}, \pi^{t},H)=0 \wedge \tau\in\{1...H\}\\
  \pi^{u}(s_t), & \text{otherwise.} 
    \end{cases}
\end{align}
\normalsize
Using this policy,~\citet{hsunguyen2023isaacs} tests on a small robot car with uncertain dynamics.

\section{Limitations and Remaining Challenges}
\label{sec:Limitations}

Hamilton-Jacobi reachability estimation has demonstrated great performance in a variety of problem formulations, even scaling up to vision-based data while providing some forms of safety guarantees. Nonetheless, there are some limitations to these approaches.

Like most learning-based approaches, acquiring the HJ reachability estimation value functions requires obtaining many samples to compute a good estimation. This may be difficult to do when trying to guarantee safety in an online framework where the number of attempts is limited. Furthermore, while recent works can guarantee convergence to the optimally safe control and value function as shown in~\citet{yu2022reachability, ganai2023iterative}, learning-based methods have issues including catastrophic forgetting~\cite{robins1995catastrophic} that make it difficult to guarantee safety within a limited number of training steps/samples.

The valid definition and formulation of the HJ reachability estimation may also be limited in the possible behaviors that it can capture. For instance, when learning the reachability formulation,~\citet{akametalu2018minimum, fisac2019bridging} had to define it in a discounted Bellman formulation. One way this was done was by defining a different optimal control problem as in~\eqref{eq:disc_optform} that incorporated discounted costs. However, the exact Bellman formulation (shown in~\eqref{eq:hjdiscsol}) to solve this had a loose gamma contraction mapping, thereby taking longer to converge to the value function solution. The other, most frequently used approach from~\eqref{eq:hjdiscsol_bridging} define a different Bellman formulation which had a tighter gamma contraction mapping -- while this formulation is a good approximation of the true Bellman formulation solution, it is not an exact reachability value function solution. Furthermore, in either case, the optimal control was redefined with discounting so the optimal control may potentially be in conflict with the true undiscounted optimal control. In other words, these HJ reachability estimation methods are limited by the learning frameworks in which they are situated that exact fundamental modifications in the HJBPDE. Thus, there remains the challenge of rectifying discrepancies between the optimal controls of the ``true" undiscounted HJBPDE versus the discounted one.

Another limitation is that the reachability value functions, especially those learned via the Bellman formulation, are rigorously defined only for deterministic dynamics or non-deterministic dynamics with known bounds~\cite{akametalu2018minimum}. Methods like those found in~\citet{yu2022safe, hsunguyen2023isaacs} that consider stochastic noise/disturbance require learning an additional model or disturbance policy. Probabilistic reachability approaches meant for stochastic environments such as~\cite{ganai2023iterative, abate2008probabilistic, chiang2015aggressive, chiang2015path, sartipizadeh2019voronoi, thorpe2021approximate} can only use HJ reachability when the cost function is redefined in a binary manner. Other stochastic reachability approaches require direct access to some form of a dynamics or control model like a probabilistic density function of the adversary's predicted control~\cite{vinod2018multiple}.

Also, as explored in~\citet{ganai2023iterative}, when the agent is outside the feasible set, the reachability value function by itself does not guarantee reentrance back into the feasible set. In particular, the control may incur a potentially infinite number of costs smaller than the maximum cost along the trajectory. Thus another challenge involves identifying how to adapt the HJBPDE into learning frameworks so it can be self-sufficient in providing such reentrance gurantees.

Finally, learning HJ reachability in a model-free manner is limited by assumptions of the online learning of the Bellman formulation. In particular, there exist novel HJ Bellman variational inequalities such as the Control Barrier Value Function variational inequality (CBFVI)~\cite{choi2021robust} whose solutions are provably both a HJ reachability value function \textit{and} a Control Barrier Function. The discrete-time solution of the CBFVI is similar to that found in~\eqref{eq:hjdiscsol} but requires $\gamma\geq 1$. However, if we want to learn the value function online via Bellman recursion, we need to ensure gamma contraction mapping which requires $\gamma\in [0,1)$. Because there is no feasible overlap in the solution space for $\gamma$, learning a Control Barrier Function with HJ reachability estimation online remains an open challenge.

\section{Future Research Directions}
\label{sec:FutureWork}
HJ reachability estimation for learning-based control is a rapidly growing field and has much more to offer. Future work includes addressing concerns about its limitations as well as extending new topics in reinforcement learning and HJ reachability.

One important domain in learned control is single lifetime reinforcement learning~\cite{chen2022you} or lifelong learning~\cite{thrun1995lifelong} in which the goal is to solve a task without resetting the environment. In the safety version of this setting, the algorithms need to be able to learn controls on the go while not terminating or entering a deadly state -- safety is a priority during exploration. Hence, there still remains the open challenge of guaranteeing safety and liveness \textit{during} the training process while interacting with the environment or from offline data so as to safely complete the task in one/few trial(s). Progress in the field of continual reinforcement learning~\cite{khetarpal2022towards, abel2024definition} can be adapted to specifically address such requirements.

Another topic to explore is HJ reachability estimation in the Koopman-Hopf framework~\cite{sharpless2023koopman}. The Hopf formula for HJ reachability analysis is an approach proposed to solve high-dimensional tasks~\cite{darbon2016algorithms, kirchner2017time, chow2019algorithm} but is limited to linear time-varying systems. Koopman theory~\cite{koopman1931hamiltonian, mezic2021koopman} is a mechanism of mapping nonlinear dynamics into some linear dynamics in a very high-dimensional latent space. There has been some work on using Koopman and reachability analysis together~\cite{kochdumper2022conformant}, but the work of~\citet{sharpless2023koopman} is novel in proposing to combine the Hopf reachability framework and Koopman theory to solve problems up to $10$-dimensions. There has been recent work improving the scalability of Koopman-based methods through learning-based mechanisms~\cite{takeishi2017learning, lusch2018deep}. This leaves room for future research tackling the challenge of further scaling Koopman-Hopf reachability analysis and applying this technique to learning-based control.

\section{Conclusion}
In this survey, we review the recent advances made in using learning-based HJ reachability estimation to reliably solve a host of challenging control tasks. While traditional HJ reachability methods have been used to safely solve complex real-world tasks (Section~\ref{sec:TradHJ_RL}), recent approaches have estimated the HJ reachability value function based on the Bellman recursive framework that learns from samples collected online (Section~\ref{subsec:Bellman_form}). With this framework, the recent literature demonstrates how we can solve various types of learning-based control tasks including standard optimal control with reinforcement learning (Section~\ref{subsec:DiscRLHJ}), reach-avoid problems (Section~\ref{sec:ReachAvoid}), and safety-constrained reinforcement learning tasks (Section~\ref{sec:ModelFreeSafeRL}). The recent also discusses works using HJ reachability estimation that address issues of robustness and generalizability to new environments of learning-based control deployed in real-world hardware. We finally discuss some challenges with using HJ reachability estimation (Section~\ref{sec:Limitations}) as well as some of its open problems that future research directions can address (Section~\ref{sec:FutureWork}). Overall, this survey serves as a primer for those interested in HJ reachability-based methods for scalable and safe learning-based control.

\newpage
\bibliographystyle{plainnat}%{unsrt}

\begin{IEEEbiography}[{\includegraphics[width=1in,height=1.1in,clip,keepaspectratio]{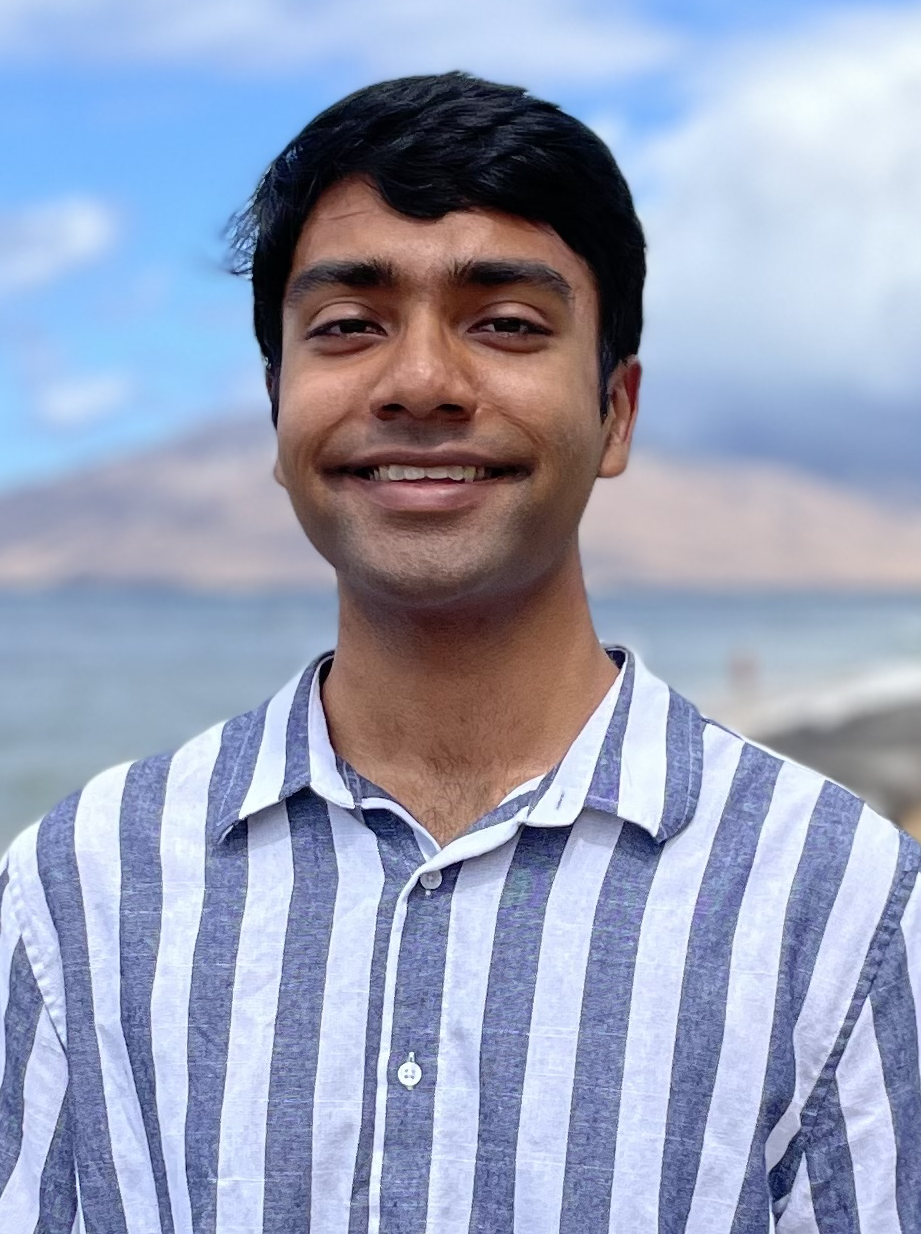}}]{Milan Ganai}{\,}(mganai@ucsd.edu) is a graduate student in the Department of Computer Science and Engineering at UC San Diego. He works on designing and creating reliable algorithms for learning-based autonomous systems. He received a B.S. in Computer Science at UC San Diego.
\end{IEEEbiography}

\begin{IEEEbiography}[{\includegraphics[width=1in,height=1.1in,clip,keepaspectratio]{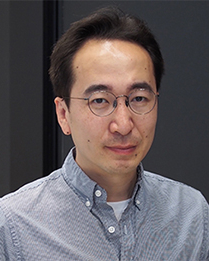}}]{Sicun Gao}{\,}(sicung@ucsd.edu) is an Associate Professor in Computer Science and Engineering at the University of California, San Diego. He works on computational methods and tools for improving automation and autonomous systems. He is a recipient of the Air Force Young Investigator Award, the NSF Career Award, and a Silver Medal for the Kurt Godel Research Prize. He received his Ph.D. from Carnegie Mellon University and was a postdoctoral researcher at CMU and MIT.
\end{IEEEbiography}

\begin{IEEEbiography}
[{\includegraphics[width=1in,height=1.25in,clip,keepaspectratio]{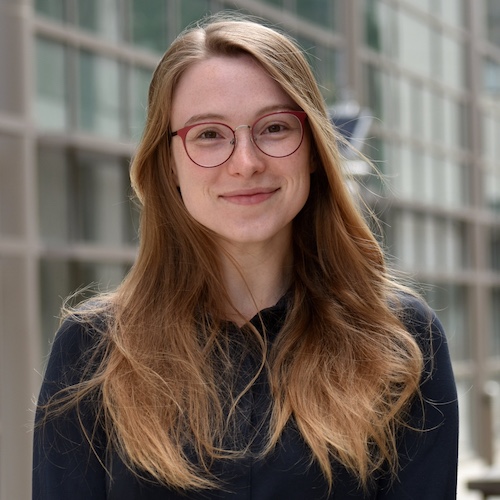}}]{Sylvia L. Herbert}{\,}(sherbert@ucsd.edu) is an Assistant Professor at UC San Diego. She received her Ph.D. from UC Berkeley in Electrical Engineering and Computer Sciences in 2020.
She works in the area of safe control for autonomous systems.
She is the recipient of an ONR Young Investigator Award, the UC Berkeley Chancellor’s Fellowship, and the Berkeley EECS Demetri Angelakos Memorial Achievement Award for Altruism. 
\end{IEEEbiography}

\end{document}